\documentclass[12pt]{iopart}

\newcommand{\PlanckMass}{M_{\rm Pl}}  
\usepackage{latexsym}
\usepackage{graphicx}
\usepackage{cite}
\begin{document}
\title[]{Cosmology of Quasi-Dilaton Massive Gravity with Non-minimal Kinetic Coupling}

\author{Sobhan Kazempour${}^1$, Amin Rezaei Akbarieh ${}^2$, Sichun Sun ${}^1$ \& Chengye Yu ${}^1$}

\address{${}^1$ School of Physics, Beijing Institute of Technology, Beijing, 100081, China}
\address{${}^2$ Faculty of Physics, University of Tabriz, Tabriz 51666-16471, Iran}
\ead{sobhan.kazempour1989@gmail.com\\ am.rezaei@tabrizu.ac.ir \\ sichunssun@gmail.com \\ chengyeyu1@hotmail.com}
\vspace{10pt}
\begin{indented}
\item[]
\end{indented}

\begin{abstract}
In this study, we introduce an extension of the quasi-dilaton massive gravity theory and derive the field equations by varying the action with respect to the metric. This extension elucidates the dynamics of the system and demonstrates how it can encompass and recover previous cosmological models through different parameter values. We present the cosmological background equations to analyze self-accelerating solutions that can explain the late-time accelerated expansion of the Universe, driven by an effective cosmological constant arising from massive gravity. Besides, we apply the quasi-dilaton massive gravity theory with non-minimal kinetic coupling to a Type Ia Supernovae (SNIa) data set to test its viability. Our findings indicate that the theory is able to account for the observed acceleration of the expansion of the universe without invoking dark energy.
In addition, we carry out a comprehensive perturbation analysis examining tensor, vector, and scalar perturbations independently. We derive the dispersion relation of gravitational waves in a Friedman-Lemaitre-Robertson-Walker (FLRW) cosmology and determine the stability conditions of the system. Such an analysis results in a sharper quasi-dilaton massive gravity theory with non-minimal kinetic coupling by ensuring the stability conditions of the system are maintained and that strong constraints on theory parameters are provided.
\end{abstract}

%
%
%
%
%

\section{\label{sec:intro}Introduction}

According to various observational evidence, such as baryon acoustic oscillations \cite{Beutler:2011hx,SDSS:2009ocz}, CMB \cite{Planck:2015fie,WMAP:2003elm}, and supernovas Ia \cite{Phillips:1993ng,SupernovaSearchTeam:1998fmf}, it is widely accepted that the Universe is undergoing an accelerated expansion phase. To elucidate this accelerated expansion and the cosmological constant problem, several approaches have been proposed \cite{Copeland:2006wr,Sotiriou:2008rp,Clifton:2011jh,Nojiri:2006ri,Bamba:2012cp}. While General Relativity is a well-established theory 
 \cite{Weinberg:1965rz}, it does not provide a satisfactory explanation for the late-time acceleration of the Universe \cite{Weinberg:1988cp,Peebles:2002gy}.

One promising solution to the cosmological conundrums lies in massive gravity theories, where gravity is mediated by a spin-2 particle with a nonzero mass \cite{deRham:2010ik,deRham:2010kj,Hinterbichler:2011tt,deRham:2014zqa,Hassan:2011hr,Hassan:2011zd}. In 1939, Fierz and Pauli introduced the first linear theory of a massive spin-2 field: Lorentz-invariant and ghost-free in a flat spacetime \cite{Fierz:1939ix}. However, this theory exhibited a discontinuity, known as the van Dam-Veltman-Zakharov (vDVZ) discontinuity, where the theory did not reduce to General Relativity in the limit of zero graviton mass \cite{vanDam:1970vg,Zakharov:1970cc}. Vainshtein resolved this issue by considering a nonlinear Fierz-Pauli action \cite{Vainshtein:1972sx}, although subsequently, Boulware and Deser identified a ghost in this nonlinear theory, known as the Boulware-Deser ghost \cite{Boulware:1972yco}. Arkani-Hamed et al. and Creminelli et al. further confirmed the instability of nonlinear massive gravity \cite{Arkani-Hamed:2002bjr,Creminelli:2005qk}.

Despite these challenges, significant progress was made in 2010 when de Rham, Gabadadze, and Tolley (dRGT) successfully constructed a fully nonlinear massive gravity theory without the Boulware-Deser ghost \cite{deRham:2010ik,deRham:2010kj}. Their theory introduced nonlinear interactions that described a massive spin-2 field in a flat spacetime. While the dRGT massive gravity can explain the accelerated expansion of the Universe without dark energy, it is only valid for an open FLRW solution and lacks stable solutions for a homogeneous and isotropic Universe \cite{DeFelice:2012mx}. Additionally, the scalar and vector perturbations in this theory face issues due to a strong coupling problem and a nonlinear ghost instability \cite{Gumrukcuoglu:2011zh}.
To address these challenges, considering additional degrees of freedom, such as an extra scalar field, has proven to be a fruitful approach. The quasi-dilaton massive gravity theory, for instance, successfully explains the accelerated expansion of the Universe in FLRW cosmology \cite{DAmico:2012hia}. However, this theory also encounters perturbations instability, leading to the development of various extensions \cite{DeFelice:2013tsa,Mukohyama:2014rca,Kazempour:2022xzy}.

In this work, we propose a new extension of the dRGT massive gravity theory by introducing a quasi-dilaton massive gravity with non-minimal kinetic coupling. Our primary motivations are twofold: firstly, to explain the late-time accelerated expansion of the Universe in a homogeneous and isotropic Universe (FLRW); and secondly, to ensure that this new theory is free from instabilities. The inclusion of a non-minimal kinetic coupling term offers several advantages, including a novel inflationary mechanism that does not rely on fine-tuned potentials \cite{Skugoreva:2013ooa}. It also provides a natural mechanism for epoch change in the early Universe \cite{Sushkov:2009hk} and alters the role of the scalar potential significantly. Power-law and Higgs-like potentials with kinetic coupling have been shown to facilitate accelerated regimes of the Universe's evolution \cite{Skugoreva:2013ooa,Sushkov:2012za,Matsumoto:2015hua}. Additionally, there have been studies exploring non-minimal massive gravity theories \cite{Bamba:2013aca,Gumrukcuoglu:2020utx}.

Numerous attempts have been made to introduce new extensions of massive gravity theories \cite{Gannouji:2013rwa,Mukohyama:2014rca,Gumrukcuoglu:2013nza,Gabadadze:2014kaa,Motohashi:2014una,Kahniashvili:2014wua,Gumrukcuoglu:2016hic,Gumrukcuoglu:2020utx,Aslmarand:2021qwn,Akbarieh:2022ovn,Kazempour:2022let,Kazempour:2022giy,Kazempour:2022xzy}, and our work contributes to this ongoing effort. We aim to present a new extension of dRGT massive gravity that can explain the accelerated expansion of the Universe in FLRW cosmology and provide stability conditions through tensor, vector, and scalar perturbations analyses. Furthermore, we will derive the dispersion relation of gravitational waves, which reveals the waveform of gravitational waves in the context of this new model.

The dRGT framework tackles essential theoretical issues like the Boulware-Deser ghost and the vDVZ discontinuity. However, it has limited use in homogeneous and isotropic cosmological models due to a lack of stable Friedmann-Lemaitre-Robertson-Walker (FLRW) solutions. This limitation drives us to expand dRGT massive gravity by incorporating a quasi-dilaton scalar field along with non-minimal kinetic coupling. By adding these components, we achieve compatibility with FLRW symmetry while maintaining the ghost-free nature of the original framework. Importantly, our approach upholds the fundamental objectives of dRGT gravity, such as deriving an effective cosmological constant from the graviton mass term $m_{g}$, all while addressing various phenomenological limitations. The introduction of non-minimal coupling alters how the scalar sector behaves, leading to stable self-accelerating solutions that correspond with observed late-time cosmic acceleration. This development not only strengthens the theoretical foundations of dRGT gravity but also expands its observational implications. It creates a testable model that distinguishes itself from current dark energy theories and scalar-tensor frameworks.

Recent progress in massive gravity theories has widened their theoretical and observational domains, resolving longstanding problems, and introducing new predictions that can be tested. For instance, Ould El Hadj showed that dRGT massive gravity modifies the emission of gravitational waves from black hole mergers and predicts different spectral features, including quasibound states and quasinormal modes that depend on parity, distinguishing itself from General Relativity \cite{OuldElHadj:2024psw}. Liu et al. demonstrate new synergies between massive gravity and the cosmological formation of the structure by showing that non-Gaussian statistics arising from redshift-space anisotropies can break degeneracies between modified gravity and massive neutrinos, an important step toward isolating signatures of massive gravity in observationally large surveys \cite{Liu:2024uxa}. Various recent numerical works studying the dRGT dynamics in spherical symmetry have also shown new ways to collapse in spherical symmetry, including naked singularity formation, and proposed a harmonic-inspired formulation to allow for full 3+1 simulations and replace some previous concerns about stability upon the nonlinear regime \cite{Albertini:2024kmf}.
Furthermore, Domcke compiled the prospects for studying massive gravity via stochastic gravitational wave backgrounds and transient signals, highlighting its contribution to bridging quantum gravity phenomenology and astrophysics \cite{Domcke:2024soc}. All these pieces of work showcase the advancement of the field in reconciling plausible observations with theoretical consistency, implying that massive gravity serves as a good framework to extend investigations of beyond-GR physics there are other similar works too \cite{Chagoya:2025hdn,Kozuszek:2024vyb,Damour:2025ucw,Chernicoff:2024dll}.

This paper is structured as follows: In Section \ref{sec:1}, we introduce the quasi-dilaton massive gravity with non-minimal kinetic coupling theory and present the background equations of motion and self-accelerating solutions. In Section \ref{sec:2}, we evaluate the quasi-dilaton massive gravity with non-minimal kinetic coupling theory using SNIa data. Section \ref{sec:3} focuses on perturbation analyses to present the stability conditions of the system. Finally, we conclude and discuss our findings in Section \ref{sec:4}.

\section{\label{sec:1}Quasi-Dilaton Massive Gravity with Non-minimal Kinetic Coupling}

In this section, we present an extended theory of quasi-dilaton dRGT massive gravity by incorporating a non-minimal kinetic coupling component. We delve into the evolution dynamics of a cosmological background, considering key elements such as the Planck mass $\PlanckMass$, the Ricci scalar $R$, a dynamic metric $g_{\mu\nu}$, and its determinant $\sqrt{-g}$. The action is given by
\begin{eqnarray}\label{Action}
S= && \frac{\PlanckMass^{2}}{2}\int d^{4} x \Bigg\{\sqrt{-g}\bigg[R +2{m}_{g}^{2}U(\mathcal{K}) - \frac{1}{\PlanckMass^{2}}(\eta G_{\mu\nu} +\omega g_{\mu\nu})\partial^{\mu}\sigma \partial^{\nu} \sigma \bigg]\Bigg\} \nonumber\\ &&  + \int d^{4}x \sqrt{-g} \mathcal{L}_{m}, \nonumber\\
\end{eqnarray}
where the $\mathcal{L}_{m}$ is the matter Lagrangian. The graviton mass arises from a multifaceted potential $U$, which can be divided into three distinct components.
\begin{equation}\label{Upotential1}
U(\mathcal{K})=U_{2}+\alpha_{3}U_{3}+\alpha_{4}U_{4},
\end{equation}
here $\alpha_3$ and $\alpha_4$ are dimensionless free parameters of the theory.
\begin{eqnarray}\label{Upotential2}
U_{2}&=&[\mathcal{K}]^{2}-[\mathcal{K}^{2}],
\nonumber\\
U_{3}&=&[\mathcal{K}]^{3}-3[\mathcal{K}][\mathcal{K}^{2}]+2[\mathcal{K}^{3}],
\nonumber\\
U_{4}&=&[\mathcal{K}]^{4}-6[\mathcal{K}]^{2}[\mathcal{K}^{2}]+8[\mathcal{K}][\mathcal{K}^{3}]+3[\mathcal{K}^{2}]^2-6[\mathcal{K}^{4}],
\end{eqnarray}
the quantity $[\mathcal{K}]$ serves as the traces of powers of the building block tensor. This tensor, denoted as $\mathcal{K}$, can be defined as
\begin{equation}\label{K}
\mathcal{K}^{\mu}_{\nu}=\delta^{\mu}_{\nu}-e^{\sigma/\PlanckMass}\sqrt{g^{\mu\alpha}f_{\alpha\nu}},
\end{equation}
where $ f_{\alpha\nu}$ is the fiducial metric, which is defined through
\begin{equation}\label{7}
f_{\alpha\nu}=\partial_{\alpha}\phi^{c}\partial_{\nu}\phi^{d}\eta_{cd},
\end{equation}
let $g^{\mu\nu}$ denote the physical metric and $\eta_{cd}$
be the Minkowski metric with $c,d= 0,1,2,3$. 
The Stueckelberg fields, denoted as $\phi^{c}$, are introduced to restore general covariance. It is important to note that $\sigma$ represents the quasi-dilaton scalar, and $\omega$ is a dimensionless free kinetic constant. We introduce a non-minimal kinetic coupling term $\eta$. It is obvious that
$G_{\mu\nu}$
 corresponds to the Einstein tensor. Furthermore, the theory exhibits invariance under global dilation transformations, implying a shift in $\sigma$ by a constant value, $\sigma_{0}$ i.e.,$\sigma\rightarrow\sigma+\sigma_{0}$.

At this stage, varying the action with respect to the metric $g_{\mu\nu}$ yields the field equations, providing insights into the dynamics of the system.
\begin{eqnarray}
\PlanckMass^{2}G_{\mu\nu}=\eta\tau_{\mu\nu}+\omega T_{\mu\nu}^{(\sigma)}+m_{g}^{2}Z_{\mu\nu},
\end{eqnarray}
where
\begin{eqnarray}
\tau_{\mu\nu}=-\frac{1}{2}\nabla_{\mu}\sigma\nabla_{\nu}\sigma R+2\nabla_{\alpha}\sigma\nabla(_{\mu}\sigma R_{\nu}^{\alpha})-\frac{1}{2}(\nabla \sigma)^{2}G_{\mu\nu}+\nabla^{\alpha}\sigma\nabla^{\beta}\sigma R_{\mu\alpha\nu\beta}\nonumber\\+\nabla_{\mu}\nabla^{\alpha}\sigma \nabla_{\nu}\nabla_{\alpha}\sigma - \nabla_{\mu}\nabla_{\nu}\sigma \Box \sigma +g_{\mu\nu}\big[-\frac{1}{2}\nabla^{\alpha}\nabla^{\beta}\sigma\nabla_{\alpha}\nabla_{\beta}\sigma +\frac{1}{2}( \Box \sigma)^{2}\nonumber\\-\nabla_{\alpha}\sigma\nabla_{\beta}\sigma R^{\alpha\beta}\big], 
\end{eqnarray}
\begin{eqnarray}
T_{\mu\nu}^{(\sigma)}=\omega \nabla_{\mu}\sigma \nabla_{\nu}\sigma -\frac{1}{2}\omega g_{\mu\nu}(\nabla \sigma)^{2},
\end{eqnarray}
\begin{eqnarray}
Z_{\mu\nu}=\mathcal{K}_{\mu\nu}-\mathcal{K}g_{\mu\nu}-\alpha\bigg\lbrace \mathcal{K}_{\mu\nu}^{2}-\mathcal{K}\mathcal{K}_{\mu\nu}+\frac{[\mathcal{K}]^{2}-[\mathcal{K}^{2}]}{2}g_{\mu\nu}\bigg\rbrace +3\beta \bigg\lbrace \mathcal{K}_{\mu\nu}^{3}\nonumber\\-\mathcal{K}\mathcal{K}_{\mu\nu}^{2}+\frac{1}{2}\mathcal{K}_{\mu\nu}\lbrace [\mathcal{K}]^{2}-[\mathcal{K}^{2}]-\frac{1}{6}g_{\mu\nu}\lbrace [\mathcal{K}]^{3}-3[\mathcal{K}][\mathcal{K}^{2}]+2[\mathcal{K}^{3}]\rbrace \bigg\rbrace. 
\end{eqnarray}
It is important to note that the resulting field equations are of second order. In the context of non-massive gravity, where $m_{g}=0$, setting $\omega =1$ and $\eta =0$ yields a trivial model without kinetic coupling. The solution to this specific case has been discussed in the literature, as referenced in \cite{Sushkov:2009hk},
\begin{eqnarray}\label{Fi-S}
a_{0}(t) = t^{1/3}, \qquad \sigma_{0}(t)=\frac{1}{2\sqrt{3\pi}}\ln{t}.
\end{eqnarray}
Also, in the case of non-massive gravity, when considering the case where $\omega =0$ and $\eta \neq 0$, resulting in a model devoid of a free kinetic term, the solution can be found in the reference \cite{Sushkov:2009hk},
\begin{eqnarray}
a(t)=t^{2/3}, \qquad \sigma(t)=\frac{t}{2\sqrt{3\pi  |\eta |}}, \quad \eta <0.
\end{eqnarray}

When $\omega =1$ and $\eta \neq 0$, the model describes an ordinary scalar field with non-minimal kinetic coupling. For sufficiently long times ($t \rightarrow \infty$), the solution to this case approximates Eq. (\ref{Fi-S}), exhibiting similar behaviour to the case of $\eta =0$. As mentioned in \cite{Saridakis:2010mf}, in the case of $\eta \neq 0$, the choice of $\omega = 1$ corresponds to a de Sitter phase, while $\omega = -1$ represents a phantom field.
For our cosmological application, we consider the FLRW Universe. The general expressions for the corresponding dynamical and fiducial metrics are as follows:
\begin{equation}\label{DMetric}
g_{\mu\nu}=diag[-N^{2},a(t)^2,a(t)^2,a(t)^2],
\end{equation}
\begin{equation}\label{FMetric}
f_{\mu\nu}=diag[-\dot{f}(t)^{2},1,1,1],
\end{equation}
in the given expression, $N$ denotes the lapse function of the dynamical metric, akin to a gauge function. The scale factor is represented by $a$, and $\dot{a}$ indicates the derivative with respect to time. It is important to note that the lapse function relates the coordinate-time $dt$ and the proper-time $d\tau$, such that $d\tau=Ndt$, as referenced in \cite{Scheel:1994yr,Christodoulakis:2013xha}. The Stueckelberg scalar function is denoted by $f(t)$, with $\phi^{0}=f(t)$ and $\frac{\partial\phi^{0}}{\partial t}=\dot{f}(t)$, as discussed in \cite{Arkani-Hamed:2002bjr}.
With these definitions and considerations, the Lagrangian for the extension of quasi-dilaton massive gravity is expressed as follows:
\begin{eqnarray}
\mathcal{L}= -3\frac{a\dot{a}^{2}}{N}\PlanckMass^{2}  +m_{g}^{2}\PlanckMass^{2}\Bigg\lbrace Na^{3}(Z-1)\times\bigg[3(Z-2)  -(Z-4)(Z-1)\alpha_{3}\nonumber\\-(Z-1)^{2}\alpha_{4}\bigg]  +\dot{f}(t)a^{4}Z(Z -1)\bigg[3  -3(Z-1)\alpha_{3}+(Z-1)^{2}\alpha_{4}\bigg]\Bigg\rbrace\nonumber\\+\frac{ a\big(\omega a^{2}N^{2}-3\eta \dot{a}^{2}\big)}{2N^{3}}\dot{\sigma}^{2} + \mathcal{L}_{m},
\end{eqnarray}
where
\begin{eqnarray}\label{XX}
Z\equiv\frac{e^{\sigma/\PlanckMass}}{a}.
\end{eqnarray}
We also can assume the matter sector $\mathcal{L}_{m}$ to consist of a perfect fluid with the energy density $\rho_{m}$ and pressure $p_{m}$.
For simplifying expressions later, one can defined,
\begin{eqnarray}
H \equiv \frac{\dot{a}}{N a}.
\end{eqnarray}

\subsection{Background Dynamics and Accelerated Expansion Solutions}\label{subsec4}

We derive the cosmological background equations for a FLRW background. 
To obtain a constraint equation, we employ the unitary gauge, setting $f(t)=t$. The significance of the unitary gauge lies in its ability to eliminate unphysical fields from the Lagrangian at the classical level through gauge transformations, as referenced in \cite{Grosse-Knetter:1992tbp}. By varying the Lagrangian with respect to $f$, we derive the constraint equation, which plays a crucial role in our analysis.
\begin{eqnarray}\label{Cons}
\frac{\delta \mathcal{L}}{\delta f} = m_{g}^{2}\PlanckMass^{2}\frac{d}{dt}\bigg[a^{4}Z(Z-1)[3-3(Z-1)\alpha_{3}+(Z-1)^{2}\alpha_{4}]\bigg]=0.
\end{eqnarray}
At this stage, we derive the Friedmann equation by varying the Lagrangian with respect to the lapse function, $N$.
\begin{eqnarray}\label{EqN}
\frac{1}{\PlanckMass^{2} a^{3}} \frac { \delta\mathcal{ L} }{ \delta N}=  3H^{2} -m_{g}^{2}(Z-1)\bigg[-3(Z-2)+(Z-4)(Z-1)\alpha_{3}\nonumber\\+(Z-1)^{2}\alpha_{4}\bigg]-\frac{1}{2 N^{2}}(\omega - 9\eta H^{2})(H N+\frac{\dot{Z}}{Z})^{2}= \rho_{m}, 
\end{eqnarray}
the equation of motion for the quasi-dilaton scalar field, $\sigma$, is given by:
\begin{eqnarray}\label{EqSig}
\frac{1}{a^{3}N}\frac{\delta\mathcal{L}}{
\delta \sigma}= \frac{1}{N^{3}}\bigg\lbrace\bigg[ -3H N \big( 2\eta \dot{H}+N(\omega -3\eta H^{2})\big)\nonumber\\+\big(\omega -3\eta H^{2}\big)\dot{N}\bigg]\dot{\sigma}+N\ddot{\sigma}\big(3\eta H^{2}-\omega \big)\bigg\rbrace + m_{g}^{2} \PlanckMass Z \bigg[ 4 r \alpha_{4} Z^{3} \nonumber\\-3 (1 + 3 r ) (\alpha_{3} + \alpha_{4} ) Z^{2} + 6 (1 + r ) ( 1+ 2 \alpha_{3} + \alpha_{4} ) Z \nonumber\\ - ( 3 + r ) ( 3 + 3\alpha_{3} + \alpha_{4} ) \bigg] =0, 
\end{eqnarray}
where
\begin{eqnarray}\label{raN}
r\equiv\frac{a}{N},
\end{eqnarray}
utilizing Eq. (\ref{XX}), we derive the following equations:
\begin{equation}
\frac{\dot{\sigma}}{N\PlanckMass}= H+\frac{\dot{Z}}{NZ}, \qquad \frac{\ddot{\sigma}}{\PlanckMass}=\frac{d}{dt}\Big(NH+\frac{\dot{Z}}{Z}\Big).
\end{equation}
The final equation of motion is obtained by varying the Lagrangian with respect to the scale factor, $a$
\begin{eqnarray}\label{Eqa}
&\frac{1}{3 a^{2} N \PlanckMass}&\frac{\delta\mathcal{L}}{
\delta a}= \nonumber\\ && 3 H^{2} + 2 \frac{\dot{H}}{N} + \frac{ (H Z N + \dot{Z})}{2 Z^{3}N^{4}} \bigg\lbrace 3 \eta H^{2} Z N^{2} ( H Z N \nonumber\\&& + \dot{Z}) + H N \dot{Z} ( \omega N + 2 \eta \dot{H} ) + H \big[ Z^{2} N^{2} ( \omega N + 6 \eta \dot{H}) \nonumber\\ && - 4 \eta N \dot{Z}^{2} + 4 \eta Z \big( N \ddot{Z} - \dot{Z} \dot{N} \big) \big] \bigg\rbrace + m_{g}^{2}\bigg\lbrace Z^{2} ( 1 + 2 \alpha_{3} \nonumber\\ && + \alpha_{4} ) - 2 Z ( 3 + 3 \alpha_{3} + \alpha_{4}) + ( 6 + 4 \alpha_{3} + \alpha_{4}) \bigg \rbrace = p_{m}.\nonumber\\
\end{eqnarray}

Due to the presence of a Bianchi identity that relates the four equations of motion, one of the equations is redundant and can be eliminated from the system.
\begin{eqnarray}
\frac{\delta S}{\delta \sigma}\dot{\sigma}+\frac{\delta S}{\delta f}\dot{f}-N\frac{d}{dt}\frac{\delta S}{\delta N}+\dot{a}\frac{\delta S}{\delta a}=0.
\end{eqnarray}

Now, we present an analysis of self-accelerating solutions as an explanation for the accelerated expansion of the Universe. To evaluate these solutions within the context of this novel extension, we integrate the Stueckelberg constraint equation (\ref {Cons}),
\begin{eqnarray}\label{Self}
Z(Z-1)\bigg[3-3(Z-1)\alpha_{3}+(Z-1)^{2}\alpha_{4}\bigg] \propto a^{-4} . \nonumber\\
\end{eqnarray}
We aim to elucidate the accelerated expansion of the Universe within the framework of quasi-dilaton massive gravity theory, incorporating a non-minimal kinetic coupling. The constant solution derived from the equation yields an effective energy density that exhibits behavior akin to a cosmological constant. As the Universe expands, the right-hand side of Eq. (\ref{Self}) diminishes proportionally to $a^{- 4}$. Consequently, over an extended period, the right-hand side of the equation decreases, resulting in the left-hand side becoming zero. This, in turn, causes $Z$ to transition into a saturated constant value, denoted as $Z_{SA}$, which satisfies the condition of equating the left-hand side of the equation (\ref{Self}) to zero.
\begin{equation}\label{Self2}
Z (Z-1)\big[3-3(Z-1)\alpha_{3}+(Z-1)^{2}\alpha_{4}\big]\bigg|_{Z=Z_{\rm SA}}=0.
\end{equation}
Equation (\ref{Self2}) presents four unique solutions. One apparent solution is $Z=0$, yet this implies that $\sigma\longrightarrow -\infty$. As this solution would be multiplied by the perturbations of the auxiliary scalars, it leads to strong coupling in both the vector and scalar sectors. Consequently, this solution must be disregarded to avoid strong coupling, as supported by the reference \cite{DAmico:2012hia,Gumrukcuoglu:2013nza}. Additionally, the solution $Z=1$ is not viable due to its implication of a vanishing cosmological constant and inconsistency, as referenced in \cite{DAmico:2012hia,Gumrukcuoglu:2013nza}. Thus, we are left with only two feasible solutions to consider further,
\begin{equation}\label{XSa}
Z_{\rm SA}^{\pm}=\frac{3\alpha_{3}+2\alpha_{4}\pm\sqrt{9\alpha_{3}^{2}-12\alpha_{4}}}{2\alpha_{4}}.
\end{equation}
By employing Equation (\ref{EqN}) in conjunction with Equation (\ref{XSa}), we derive the modified Friedmann equation as follows:
\begin{eqnarray}\label{EqFr}
 \bigg[ 3 - \frac{1}{2}\big( \omega - 9 \eta H^{2} \big) \bigg] H^{2} =  \Lambda_{\rm SA}^{\pm} + \rho_{m}, \nonumber\\
\end{eqnarray}
where the effective cosmological constant, as derived from the massive component of the action, is incorporated within the modified Friedmann equation,
\begin{eqnarray}
\Lambda_{\rm SA}^{\pm}\equiv  m_{g}^{2}(Z_{\rm SA}^{\pm}- 1)\Big[ 
6-3Z_{\rm SA}^{\pm} +(Z_{\rm SA}^{\pm} - 4)(Z_{\rm SA}^{\pm} - 1)\alpha_{3} +(Z_{\rm SA}^{\pm} - 1)^{2}\alpha_{4}\Big]. \nonumber\\
\end{eqnarray}
To accurately represent the relationship, the above equation should be reformulated by considering the expression given in Equation (\ref{XSa}),
\begin{eqnarray}
\Lambda_{\rm SA}^{\pm}= && \frac{3m^{2}_{g}}{2\alpha^{3}_{4}}\bigg[9\alpha^{4}_{3}\pm 3\alpha^{3}_{3}\sqrt{9\alpha^{2}_{3}-12\alpha_{4}}-18\alpha^{2}_{3}\alpha_{4}\mp 4\alpha_{3}\alpha_{4}\sqrt{9\alpha^{2}_{3}-12\alpha_{4}} \nonumber\\ && +6\alpha^{2}_{4}\bigg]. 
\end{eqnarray}
Consequently, the expression for $H^{2}$ is provided by equation (\ref{EqFr}),
\begin{eqnarray}\label{H2}
H^{2}= \frac{\omega - 6 \pm 2 \sqrt{( 3 - \frac{\omega }{2})^{2} + 18 \eta \big( \Lambda_{\rm SA}^{\pm} + \rho_{m}} \big) }{18 \eta}.
\end{eqnarray}
In conclusion to this section, we derive the value of $r_{\rm SA}$ utilizing Equation (\ref{EqSig}),
\begin{eqnarray}\label{rSA}
r_{\rm SA}= 1 + \frac{H^{2}\bigg( \omega - 3\eta H^{2} \bigg)}{m_{g}^{2}(Z_{\rm SA}^{\pm})^{2}(Z_{\rm SA}^{\pm}\alpha_{3}-\alpha_{3}-2)}.
\end{eqnarray}
Indeed, we employed the Stuckelberg equation (\ref{Self}) to eliminate $\alpha_{4}$. The resulting equation elucidates a self-accelerating Universe devoid of strong coupling. Thus, we have demonstrated that this theoretical framework encompasses self-accelerating solutions characterized by an effective cosmological constant.

\begin{figure}
\centering
\includegraphics[width=7cm]{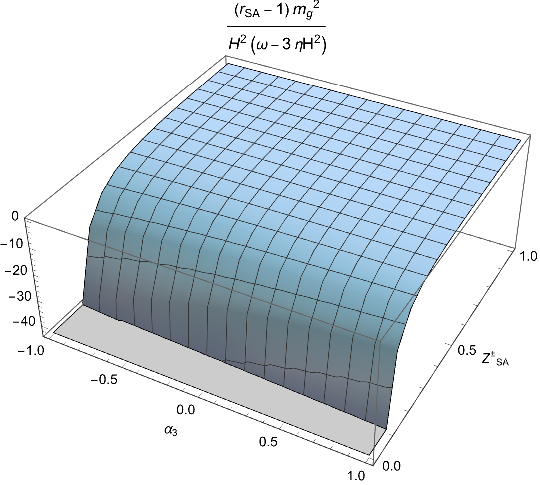}
\includegraphics[width=0.8cm]{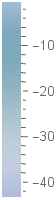}
\caption[figs]
{Plot of the $\frac{(r_{\rm SA}-1)m_{g}^{2}}{H^{2} (\omega - 3 \eta H^{2}}$ using Eq. (\ref{rSA}), by considering $0 < Z_{\rm SA}^{\pm} < 1$. The excluded regions are illustrated in grey color.}
\label{fig1}
\end{figure}
\begin{figure}
\centering
\includegraphics[width=7cm]{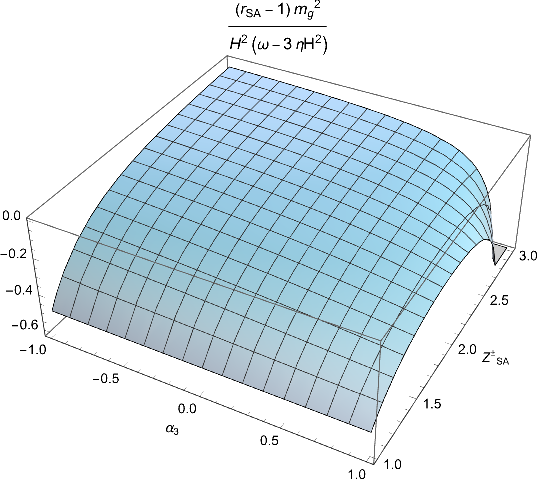}
\includegraphics[width=0.8cm]{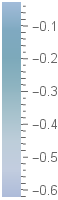}
\caption[figs]
{Plot of the $\frac{(r_{\rm SA}-1)m_{g}^{2}}{H^{2} (\omega - 3 \eta H^{2}}$ using Eq. (\ref{rSA}), by considering $ Z_{\rm SA}^{\pm} > 1$. The excluded regions are illustrated in grey color.}
\label{fig2}
\end{figure}

In Figures \ref{fig1} and \ref{fig2}, we present a graphical representation of equation (\ref{rSA}), which illustrates the acceptable ranges for the parameters $\frac{(r_{\rm SA}-1)m_{g}^{2}}{H^{2} (\omega - 3 \eta H^{2}}$, $\alpha_{3}$ and $Z_{\rm SA}^{\pm}$.

\section{Cosmological Test}\label{sec:2}

The exploration of Type Ia supernovae provided compelling evidence for the accelerated expansion of the universe \cite{Copeland:2006wr,Frieman:2008sn,Perlmutter:2003kf,Yang:2019fjt}. Here, we utilize the Union2 supernovae Ia dataset, comprising 557 SNIa events \cite{Amanullah:2010vv}, to investigate the quasi-dilaton massive gravity with non-minimal kinetic coupling theory. The observational data from the SNIa dataset will be presented in terms $\mu_{\rm obs}$, enabling a direct comparison with the theoretical predictions of the model under scrutiny.

\begin{eqnarray}
\mu_{\rm th}(z_{i})=5\log_{10}D_{L}(z_{i})+\mu_{0},
\end{eqnarray}
it is important to note that $\mu_{0}=42.38 - 5 \log_{10} \mathcal{H}$, where $\mathcal{H}$ represents the Hubble constant $H_{0}$ in units of $100 \, {\rm km/s/Mpc}$. This relationship is essential for establishing the connection between the observed magnitude and the luminosity distance, which can be expressed as follows:
\begin{eqnarray}
D_{L}(z)=(1+z)\int_{0}^{z}\frac{dx}{E(x;q)},
\end{eqnarray}
furthermore, it is evident that $E=H/H_{0}$ and $q$ are the model parameters. It is important to note that $X^{2}$ is defined as follows:
\begin{equation}\label{3.3}
X_{\mu}^{2}(q)=\sum_{i}\frac{[\mu_{\rm obs}(z_{i})-\mu_{\rm th}(z_{i})]^{2}}{\sigma^{2}(z_{i})},
\end{equation}
here, $\sigma$ corresponds to the $1\sigma$ error, and the parameter $\mu_{0}$ is a nuisance parameter independent of the data points. To minimize $X_{\mu}^{2}$ in equation (\ref{3.3}), we expand it with respect $\mu_{0}$ \cite{Nesseris:2005ur,DiPietro:2002cz}.
\begin{equation}\label{3.4}
X_{\mu}^{2}(q)=\tilde{T}-2\mu_{0}\tilde{Y}+\mu_{0}^{2}\tilde{Z},
\end{equation}
where
\begin{eqnarray}
&&\tilde{T}(q)=\sum_{i}\frac{[\mu_{\rm obs}(z_{i})-\mu_{\rm th}(z_{i};\mu_{0}=0,q)]^{2}}{\sigma_{\mu_{\rm obs}}^{2}(z_{i})}, \nonumber\\
&&\tilde{Y}(q)=\sum_{i}\frac{\mu_{\rm obs}(z_{i})-\mu_{\rm th}(z_{i};\mu_{0}=0,q)}{\sigma_{\mu_{\rm obs}}^{2}(z_{i})}, \nonumber\\
&&\tilde{Z}=\sum_{i}\frac{1}{\sigma_{\mu_{\rm obs}}^{2}(z_{i})}.
\end{eqnarray}
It should be explained that for $\mu_{0}=\frac{\tilde{Y}}{\tilde{Z}}$, equation (\ref{3.4}) has a minimum at
\begin{equation}
\tilde{X}_{\mu}^{2}(q)=\tilde{T}(q)-\frac{\tilde{Y}^{2}(q)}{\tilde{Z}}.
\end{equation}
As it is clear that $X_{\mu, {\rm min }}^{2}=\tilde{X}_{\mu, {\rm min}}^{2}$ we can consider minimizing $\tilde{X}_{\mu}^{2}$, which is independent of $\mu_{0}$. It is important to note that the best-fit model parameter is determined by minimizing $X^{2}=\tilde{X}_{\mu}^{2}$. At the same time, we know that the corresponding $\mathcal{H}$ is dependent on $\mu_{0}=\frac{\tilde{Y}}{\tilde{Z}}$ for the best-fit parameter.

By considering equation (\ref{H2}) and applying the change of variables $a=\frac{1}{1+z}$ and $\frac{d}{dt}=-H(z+1)\frac{d}{dz}$, we can express the dimensionless Hubble parameter for this case. In our analysis, we focus on the late-time accelerated expansion of the Universe, utilizing the Taylor expansion around $z = 1$. Our model is specifically designed to fit low redshift data, with a focus on values around $z \leq 1$. Note that in our calculations the matter density has been fixed. Consequently, the solution to the asymptotic state at small redshifts should be provided as follows:
\begin{eqnarray}
H(z) \sim H_{0} + \frac{\Theta_{\pm}}{H_{0}^{3}}z - \frac{1}{2} (\frac{\Theta_{\pm}^{2}}{H_{0}^{7}} + \frac{\Theta_{\pm}}{H_{0}^{3}} - \frac{H_{0}}{2} ) z^{2} +.....,   \nonumber\\ 
\end{eqnarray}
where $\Theta_{\pm}$ is defined as
\begin{eqnarray}
\Theta_{\pm} = \frac{54 \eta \Omega}{\varsigma \big( \mp 6 \pm \omega + \varsigma \big) },  \quad  \varsigma = \sqrt{36 +72 \eta \Lambda_{\rm SA}^{\pm}-12 \omega + \omega^{2}+ 72\eta \Omega}.
\end{eqnarray}
it is important to note that $H_{0}$ represents the Hubble parameter at the present time.
Therefore, the dimensionless Hubble parameter can be expressed as follows:
\begin{eqnarray}
E=\frac{H(z)}{H_{0}}= 1 + F z + ( 2 F -F^{2}) z^{2} + ...... \nonumber\\
\end{eqnarray}
where
\begin{eqnarray}
F = \frac{\Theta_{\pm}}{H_{0}^{4}}.
\end{eqnarray}

In this step, we present plots of $X^{2}$ and likelihoods as functions of the parameter $F$. Based on our calculations, the best-fit yields a value of $X_{\rm min}^{2}=543.28$, and the corresponding best-fit parameter is,
\begin{eqnarray}
F=& 0.2991^{+0.0225}_{-0.0218}, \quad\quad \mbox{with 1$\sigma$ uncertainty}, \\
F=& 0.2991^{+0.0460}_{-0.0428}, \quad\quad \mbox{with 2$\sigma$ uncertainty}.
\end{eqnarray}
Consequently, it is noteworthy to mention that the best-fit value for the quasi-dilaton massive gravity with non-minimal kinetic coupling theory corresponds to $\mathcal{H}=0.70248$.

\begin{figure}
\centering
\includegraphics[width=8cm]{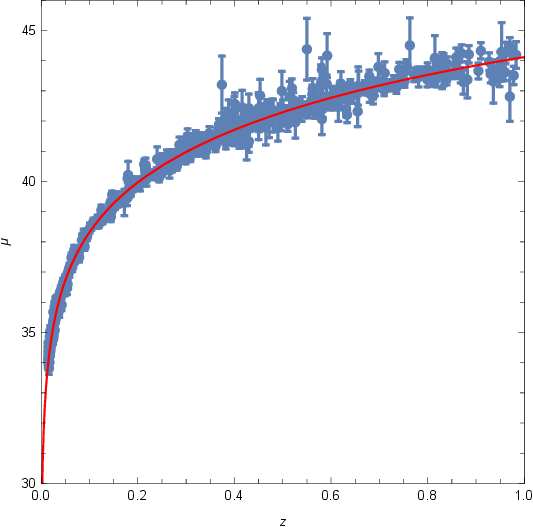}
\caption[figs]
{The distance modulus diagram for the best fit (red solid line) of the parameter of the quasi-dilaton massive gravity with non-minimal kinetic coupling theory in comparison with the 557 Union2 SNIa data points (blue dots).}
\label{Nfig1}
\end{figure}

\begin{figure}
\centering
\includegraphics[width=8cm]{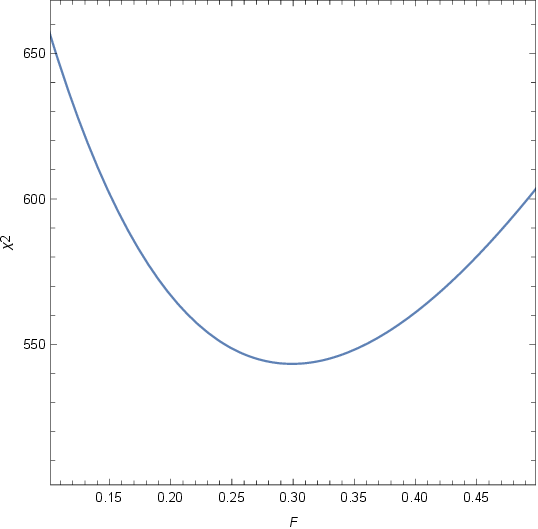}
\caption[figs]
{The $X^{2}$ functions of parameter $F$ for the quasi-dilaton massive gravity with non-minimal kinetic coupling theory.}
\label{Nfig2}
\end{figure}

\begin{figure}
\centering
\includegraphics[width=8cm]{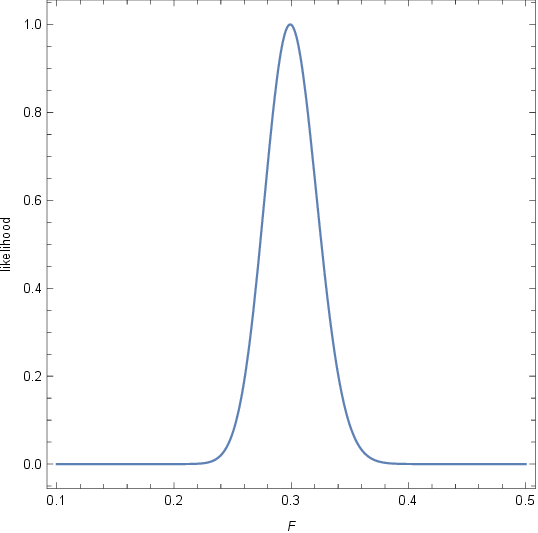}
\caption[figs]
{The likelihood as functions of parameter $F$ for the quasi-dilaton massive gravity with non-minimal kinetic coupling theory.}
\label{Nfig5}
\end{figure}
According to Figures \ref{Nfig1} to \ref{Nfig5}, the result of fitting the quasi-dilaton massive gravity with non-minimal kinetic coupling theory to the cosmological data provides us with the optimal value for the model parameter.

To evaluate the validity of this theory, we utilize the Union2 supernovae Ia dataset, consisting of $557$ SNIa events. By employing the Bayesian statistics method, we constrain the model parameters based on the SNIa data. Specifically, we generate plots of $X^{2}$ and the likelihood function as functions of the parameter $F$, allowing us to determine the best-fit value by minimizing $X^{2}$.

Our analysis yields a best-fit value of $F=0.2991$ corresponding to a minimum $X^{2}$ value of $543.28$. Using this optimal parameter value, we calculate the best-fit value of $\mathcal{H}$ for the theory, obtaining $\mathcal{H} = 0.70248$.

Figure \ref{Nfig1} illustrates the distance modulus diagram, depicting the relationship between the distance modulus $\mu$ and the redshift $z$, along with the best-fit curve for the quasi-dilaton massive gravity with non-minimal kinetic coupling theory based on supernovae Ia data.

\section{Perturbation Analyses}\label{sec:3}

We aim to present the perturbation analyses to show the stability of the solutions.

As a preliminary step, we seek to obtain the action up to the second order in perturbations. To achieve this, we consider small fluctuations, denoted as $\delta g_{\mu\nu}$, and perform an expansion of the physical metric relative to a background solution, represented as $g_{\mu\nu}^{(0)}$,
\begin{equation}
g_{\mu\nu}=g_{\mu\nu}^{(0)}+\delta g_{\mu\nu}.
\end{equation}
Pay attention that the metric perturbations could be divided into three parts,
namely tensor, vector, and scalar perturbations. Thus, we have
\begin{eqnarray}
\delta g_{00}=&&-2N^{2} \Phi, \nonumber\\
\delta g_{0i}=&&Na(B_{i}+\partial_{i}B), \nonumber\\
\delta g_{ij}=&&a^{2}\bigg[h_{ij}+\frac{1}{2}(\partial_{i}E_{j}+\partial_{j}E_{i})+2\delta_{ij}\Psi +\big(\partial_{i}\partial_{j} -\frac{1}{3}\delta_{ij}\partial_{l}\partial^{l}\big)E\bigg],
\end{eqnarray}
Moreover, there are these conditions $\delta^{ij}h_{ij}=\partial^{i}h_{ij}=\partial^{i}E_{i}=\partial^{i}B_{i}=0$ for scalar, vector, and tensor perturbations. These conditions indicate that the tensor perturbations are transverse and traceless. All perturbations are compliant with spatial rotation transformations and are functions of time and space.

One can perturb the scalar field $\sigma$ as follows
\begin{equation}
\sigma =\sigma^{(0)} + \PlanckMass \delta\sigma.
\end{equation}

It is important to note that all calculations are executed in the unitary gauge, eliminating the need to consider gauge-invariant combinations explicitly. The spatial indices on perturbations are raised and lowered using $\delta^{ij}$ and $\delta_{ij}$, respectively. Furthermore, to simplify the analysis, we express the expanded action in the Fourier domain using plane waves, substituting $\vec{\nabla}^{2}\rightarrow -k^{2}$ and $d^{3}x\rightarrow d^{3}k$.

\subsection{Tensor}\label{tensor}

In this subsection, we undertake a tensor perturbation analysis to derive the dispersion relation of gravitational waves.
Let us proceed to examine tensor perturbations,

\begin{eqnarray}
\delta g_{ij}=a^{2}h_{ij}^{TT},
\end{eqnarray}
where
\begin{equation}
\partial^{i}h_{ij}=0, \qquad g^{ij}h_{ij}=0.
\end{equation}
The tensor-perturbed action, up to the second order, can be derived independently for each component of the total action. The contribution arising from the gravity part of the perturbed action, expressed in quadratic order, is as follows:
\begin{eqnarray}
S^{(2)}_{\rm gravity}=\frac{\PlanckMass^{2}}{8}\int d^{3}k \, dt \, a^{3}N \bigg[ \frac{\dot{h}_{ij}\dot{h}^{ij}}{N^{2}} -\Big(\frac{k^{2}}{a^{2}}+4\frac{\dot{H}}{N}+6H^{2} \Big)h^{ij}h_{ij}\bigg] - \frac{2}{\PlanckMass^{2}} \mathcal{L}_{m}.  \nonumber\\
\end{eqnarray}
\\
The second-order component of the massive gravity sector within the perturbed action can be expressed as follows:
\begin{eqnarray}
S^{(2)}_{\rm massive}= \frac{\PlanckMass^{2}}{8}\int d^{3}k \, dt \, a^{3}Nm_{g}^{2}\bigg[(\alpha_{3}+\alpha_{4}) r Z^{3} - (1+2\alpha_{3}+\alpha_{4})(1 \nonumber\\ +3r)Z^{2}+(3+3\alpha_{3}+\alpha_{4})(3+2r)Z-2(6+4\alpha_{3}+\alpha_{4})\bigg]h^{ij}h_{ij}.
\end{eqnarray}
Additionally, we present the quasi-dilaton with non-minimal kinetic coupling part of the perturbed action, expressed in quadratic order:
\begin{eqnarray}
S^{(2)}_{\rm QdNm}= && -\frac{1}{8 N^{4}}\int d^{3}k \, dt \, a^{3}N\bigg[\eta\dot{\sigma}^{2}\big(\frac{3\ddot{a}}{a}-\frac{k^{2}N^{2}}{2a^{2}}-3(\frac{\dot{a}^{2}}{a^{2}}\nonumber\\ && + \frac{\ddot{a}}{a} )\big)+\omega N^{2}\dot{\sigma}^{2}\bigg]h^{ij}h_{ij}. 
\end{eqnarray}
By summing the second-order components of the perturbed actions, namely $S^{(2)}_{\rm gravity}$,  $S^{(2)}_{\rm massive}$, and $S^{(2)}_{\rm QdNm}$, we derive the total action up to the second order for tensor perturbations:

\begin{eqnarray}
S^{(2)}_{\rm total}= \frac{\PlanckMass^{2}}{8}\int d^{3}k \, dt \, a^{3}N\bigg\lbrace \frac{\dot{h}_{ij}\dot{h}^{ij}}{N^{2}}-\Big(\frac{k^{2}}{a^{2}} +M_{\rm GW}^{2}\Big)h^{ij}h_{ij}\bigg\rbrace .\nonumber\\
\end{eqnarray}
Consequently, we derive the dispersion relation for gravitational waves as follows:
\begin{eqnarray}\label{MGW}
M_{\rm GW}^{2}= &&  4\frac{\dot{H}}{N}+6H^{2} + \frac{1}{\PlanckMass^{2}N^{4}} \bigg[ \eta  \dot{\sigma} \big( 3 \frac{\ddot{a}}{a} - \frac{k^{2}}{2 a^{2}}N^{2} - 3 ( \frac{\dot{a}^{2}}{a^{2}} + \frac{\ddot{a}}{a} ) \big) \nonumber\\ &&  + \omega N^{2} \dot{\sigma}^{2} \bigg] + \frac{2}{\PlanckMass^{2}}p_{m} + \gamma, 
\end{eqnarray}
where
\begin{eqnarray}\label{gamma}
\gamma = \frac{m_{g}^{2}}{Z_{\rm SA}^{\pm}-1}\bigg[ r_{SA} Z_{\rm SA}^{\pm 4}\alpha_{3} + 2 ( 3 + \alpha_{3} ) - 2 Z_{\rm SA}^{\pm} ( 9 + 4 \alpha_{3} ) + Z_{\rm SA}^{\pm 2} ( 10 + 8 \alpha_{3} \nonumber\\ + 3 r_{SA} (2 + \alpha_{3} ) )  - Z_{\rm SA}^{\pm3} ( 1 + 2 \alpha_{3} + r_{SA} (3 + 4 \alpha_{3}) ) \bigg].
\end{eqnarray}

We employed Equation (\ref{XSa}) to eliminate $\alpha_{4}$ and considered the value of $r_{SA}$.

In this section, the main point is that stability conditions would be met if the square of the mass of gravitational waves is positive, ensuring the stability of long-wavelength gravitational waves, i.e., $M_{\rm GW}^{2} > 0$.  Conversely, a negative value indicates a tachyonic nature. In the context of tachyonic instability, it is worth mentioning that the mass of the tachyon is on the order of the Hubble scale, implying that the instability would require a duration comparable to the age of the Universe to manifest significantly.

\begin{figure}
\centering
\includegraphics[width=7cm]{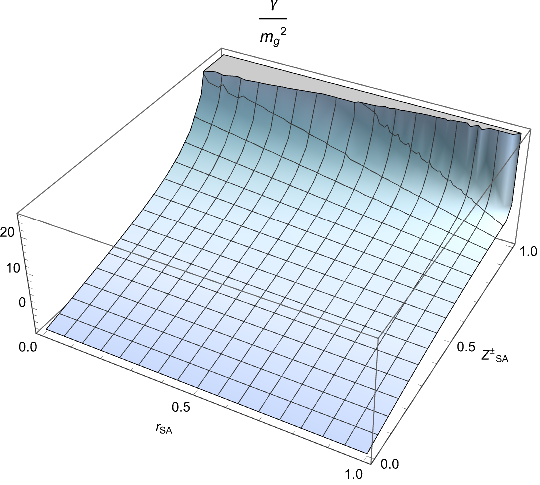}
\includegraphics[width=0.8cm]{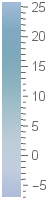}
\caption[figs]
{Plot of the $\frac{\gamma}{m_{g}^{2}}$ using Eq. (\ref{gamma}), by considering $0 < Z_{\rm SA}^{\pm} < 1$, and $\alpha_{3}=0.5$. The excluded regions are illustrated in grey color.}
\label{Tfig1}
\end{figure}
\begin{figure}
\centering
\includegraphics[width=7cm]{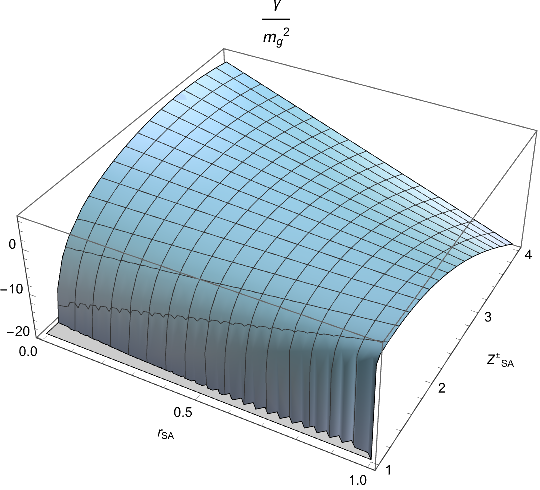}
\includegraphics[width=0.8cm]{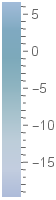}
\caption[figs]
{Plot of the $\frac{\gamma}{m_{g}^{2}}$ using Eq. (\ref{gamma}), by considering $ Z_{\rm SA}^{\pm} > 1$, and $\alpha_{3}=0.5$. The excluded regions are illustrated in grey color.}
\label{Tfig2}
\end{figure}

Figures (\ref{Tfig1}) and (\ref{Tfig2}) present a graphical representation of the possible values for the parameter $\frac{\gamma}{m_{g}^{2}}$ as a function of $\alpha_{3}$, $r_{SA}$ and $Z_{\rm SA}^{\pm}$. Notably, the parameter $\gamma$ corresponds to the massive gravity contribution to the dispersion relation of gravitational waves, and its behavior is crucial in establishing the stability of quasi-dilaton massive gravity theories with non-minimal kinetic coupling.

The second finding of this section is the modified dispersion relations of gravitational waves, presented in Equation (\ref{MGW}). This equation describes the propagation of gravitational perturbations within the context of FLRW cosmology in the framework of quasi-dilaton massive gravity with non-minimal kinetic coupling. In theory, this propagation can be experimentally verified through observations of cosmological events, particularly through gravitational wave detection.

The modifications introduced in this equation will contribute additional factors to the phase evolution of the gravitational waveform \cite{Will:1997bb,Mirshekari:2011yq}. Sensitive matched-filtering techniques employed in data analysis will be crucial for detecting these subtle effects.

The detection of gravitational waves from a merging binary black hole (GW150914) by the LIGO/Virgo Collaboration marked a pivotal moment in the renewed interest in testing the mass of gravitons \cite{LIGOScientific:2016lio,LIGOScientific:2019fpa,LIGOScientific:2020tif,Shao:2020shv}. While the corresponding Compton wavelength is significantly smaller than the Hubble scale, limiting its immediate implications for modified cosmology, prospects warrant continued exploration.

By employing more sensitive space-based gravitational-wave detectors, we can further scrutinize this fundamental aspect of gravitation by studying gravitational events at varying wavelengths. This is particularly important given the complementary nature of our theory-specific analysis in quasi-dilaton massive gravity with non-minimal kinetic coupling to the work of Nishizawa and Arai, who investigated modified propagation of gravitational waves in a parametrized framework, considering factors such as a running Planck mass and modified speed for gravitation \cite{Nishizawa:2017nef,Arai:2017hxj,Nishizawa:2019rra}. Our findings will serve as a valuable target for subsequent detailed investigations in this field.

\subsection{Vector}

Here, we provide a comprehensive analysis of the vector perturbations in quasi-dilaton massive gravity with non-minimal kinetic coupling theory. The main incentive for this work is to look for the conditions of the theory to stay stable and not exhibit instabilities.

One can consider the vector perturbations,
\begin{eqnarray}\label{Bi}
B_{i}=\frac{a \big( k^{4} (r^{2}-1)^{2} + 12 a^{4} H^{6} \eta \omega \big) }{2\big[ k^{2} (r^{2} - 1) - 6 a^{2} H^{4} \eta \big] \big[ k^{2} (r^{2} -1) + 2 a^{2} H^{2} \omega \big]}\frac{\dot{E}_{i}}{N}.
\end{eqnarray}
The field $B_{i}$, since it is non-dynamical, can be included in the action as an auxiliary field. As a result, this method leads to the presence of just one propagating vector in the system.

\begin{eqnarray}\label{AVc}
S_{\rm vector}^{(2)}=\frac{\PlanckMass^{2}}{8}\int d^{3}k \, dt \, a^{3}N 
\bigg(\frac{\chi}{N^{2}} |\dot{E}_{i}|^{2} -\frac{k^{2}}{2}M_{\rm GW}^{2}|E_{i}|^{2}\bigg),\nonumber\\
\end{eqnarray}
where
\begin{eqnarray}\label{Betta}
\chi =\frac{k^{2}}{2}\bigg(1+\frac{k^{2}(r^{2}-1) \big( 3 H^{2} \eta - \omega \big)}{ 6 a^{2} H^{4} \eta \omega} \bigg).
\end{eqnarray}
It can be found that there are two cases, in the first one, if we have $\frac{(r^{2}-1) \big( 3 H^{2} \eta - \omega \big)}{ 6 a^{2} H^{4} \eta \omega}\geq 0$, we do not need the critical momentum scale.
However, in the second one, if we have $\frac{(r^{2}-1) \big( 3 H^{2} \eta - \omega \big)}{ 6 a^{2} H^{4} \eta \omega } <0$, we should have a critical momentum scale $k_{c}= a H^{2}\sqrt{\frac{6 \eta \omega}{(1 - r^{2}) (3 H^{2}\eta - \omega)}}$ to avoid a ghost. To maintain the stability of the system, it is crucial for the physical critical momentum scale, $k_{c}/a$, to exceed the ultraviolet cutoff scale of the effective field theory, $\Lambda_{UV}$. This requirement can be stated as follows:
\begin{eqnarray}\label{39}
\Lambda_{UV}^{2} \leq\frac{6 H^{4}\eta \omega}{(1 - r^{2}) (3 H^{2}\eta - \omega)} \quad {\rm if} \quad \frac{(r^{2}-1) \big( 3 H^{2} \eta - \omega \big)}{ 6 a^{2} H^{4} \eta \omega} <0, \nonumber\\
\end{eqnarray}

Subsequently, we will examine the canonically normalized fields to identify potential instabilities in the vector modes, as detailed below:
\begin{eqnarray}\label{Khi}
\Xi_{i}=\frac{\chi E_{i}}{2}.
\end{eqnarray}
We have substituted Eq. (\ref{Khi}) into Eq. (\ref{AVc}), as shown below:
\begin{eqnarray}
S=\frac{1}{2}\int d^{3}k \, dt \, a^{3}N \bigg(\frac{|\dot{\Xi_{i}}|^{2}}{N^{2}}-c_{V}^{2}|\Xi_{i}|^{2}\bigg).
\end{eqnarray}
The speed of sound for vector modes can be expressed as:
\begin{eqnarray}\label{c_V}
c_{V}^{2}=M_{\rm GW}^{2}(1+j^{2})-\frac{H^{2}j^{2}(1+4 j^{2})}{(1+ j^{2})^{2}},
\end{eqnarray}
here the dimensionless quantity is considered:
\begin{eqnarray}
j^{2}\equiv \frac{k^{2}(r^{2}-1) \big( 3 H^{2} \eta - \omega \big)}{6 a^{2}H^{4}\eta \omega}.
\end{eqnarray}
We examine the circumstances that may result in instability in the system. The vector modes may be susceptible to a gradient instability originating from the first term in Eq. (\ref{c_V}), under the conditions $M_{\rm GW}^{2}<0$ and $j^{2}>0$. If we consider all physical momenta below the UV cut-off $\Lambda_{UV}$, thus, a growth rate of instability should be lower than the cosmological scale. So, we have
\begin{eqnarray}
&& \Lambda_{UV}^{2} \leq\frac{6 H^{4}\eta \omega}{(1 - r^{2}) (3 H^{2}\eta - \omega)}, \quad  {\rm if} \quad \frac{(r^{2}-1) \big( 3 H^{2} \eta - \omega \big)}{ 6 a^{2} H^{4} \eta \omega} > 0 \nonumber\\ && \quad {\rm and} \quad M_{\rm GW}^{2}<0. \nonumber\\
\end{eqnarray}
By examining the second part of Eq. (\ref{c_V}) more closely, there are two possibilities. In the first position, with $j^{2}>0$, there is no instability faster than the Hubble expansion, and hence the system is stable. In the second position, with $j^{2}<0$, it is necessary to satisfy the no-ghost condition of Eq. (\ref{39}) in order to maintain stability. Furthermore, for the condition of stability to be met, we must require the inequality $|j^{2}| \leq (k^{2}/a^{2})/\Lambda_{UV}^{2}$ to be satisfied. Under that condition, Eq. (5) is stable with respect to its second part. To ensure the stability of vector modes globally, it is again necessary to check that $c_{V}^{2}>0$, and furthermore, the condition for the mass squared term in the gravitational wave dispersion relation must be positive to ensure stability $M_{\rm GW}^{2}>0$, which was previously discussed in Subsection (\ref{tensor}).

\subsection{Scalar}

In this subsection, we provide an in-depth analysis of the scalar perturbations in quasi-dilaton massive gravity with non-minimal kinetic coupling. The analyses will be directed toward understanding the stability of the system.

One can begin with the action quadratic in scalar perturbations,
\begin{eqnarray}
\delta g_{00}=&&-2N^{2} \Phi, \nonumber\\
\delta g_{0i}=&&N\,a\,\partial_{i}B, \nonumber\\
\delta g_{ij}=&&a^{2}\bigg[2\delta_{ij}\Psi +\big(\partial_{i}\partial_{j} -\frac{1}{3}\delta_{ij}\partial_{l}\partial^{l}\big)E\bigg],
\end{eqnarray}
\begin{equation}
\sigma =\sigma^{(0)} + \PlanckMass \delta\sigma.
\end{equation}
Since the perturbations $\Phi$ and $B$ do not include time derivatives, these fields can be used as auxiliary fields using their equations of motion.
\begin{eqnarray}
B= \frac{(r^{2}-1) \bigg[ 6 H N \bigg( \Phi \omega - 3 H^{2} \eta ( \Phi - \delta\sigma \omega ) \bigg) + \bigg( 3 H^{2} \eta - \omega \bigg) \bigg( k^{2} \dot{E} + 6 \dot{\Psi} \bigg) \bigg]}{9 a H^{4}N \eta \omega}, \nonumber\\
\end{eqnarray}
\begin{eqnarray}
\Phi = - \frac{k^{2} \dot{E} + 6 \dot{\Psi}}{6 H N} + \bigg[12 k^{2} (r^{2}-1) - 3 a^{2}H^{2} \omega (\omega -6)\bigg]^{-1} \Bigg\lbrace k^{4} \omega^{2} \delta\sigma (3 + E ) \nonumber\\ \bigg( 2 k^{2} (r^{2}-1) - \frac{3 a^{2}H^{2}\omega}{r - 1} \bigg) + 3 \Psi\omega \bigg( 2 k^{2} + \frac{3 a^{2}H^{2}\omega}{r - 1} \bigg)  - \bigg( \frac{3 a^{2}H \omega \big( \omega \delta\dot{\sigma} - 6 \dot{\Psi} \big)}{N} \bigg) \nonumber\\ + \bigg( \frac{2 k^{2} (r^{2} -1 ) (k^{2}\dot{E} + 6 \dot{\Psi})}{HN} \bigg) \Bigg\rbrace + \bigg[ 12 k^{2} (r^{2} -1 ) - 27 a^{2} H^{4} \eta (2 + H^{2}\eta) \bigg]^{-1} \nonumber\\ \Bigg\lbrace 9 H^{2} \eta (r-1)^{-1}\bigg( H^{2}k^{4} \delta\sigma ( 3 + E )\eta \bigg[ 2 k^{2} (r -1)^{2} (r + 1) + 9 a^{2} H^{4} \eta \bigg] \nonumber\\ - \bigg[ 2 k^{2} (r-1) - 9 a^{2} H^{4} \eta \bigg] \Psi \bigg) + H^{-1}N^{-1}  \bigg[-27 a^{2} H^{4} \eta \bigg( H^{2} \eta \delta\dot{\sigma} + 2\dot{\Psi} \bigg) \nonumber\\ + 2 k^{2} (r^{2}-1) \bigg( k^{2}\dot{E} + 6 \dot{\Psi}  \bigg) \bigg] \Bigg\rbrace.
\end{eqnarray}
It is worth noting that when these equations are substituted into the action, we obtain an action that includes the three fields $E$, $\Psi$, and $\delta\sigma$.
In parallel, we identify another non-dynamical combination to eliminate the sixth degree of freedom, as follows,
\begin{eqnarray}
\tilde{\Psi}= \frac{1}{\sqrt{2}}(\Psi +\delta\sigma).
\end{eqnarray}
Furthermore, we define an orthogonal combination as,
\begin{eqnarray}
\tilde{\delta\sigma}=\frac{1}{\sqrt{2}k^{2}}(\Psi -\delta\sigma).
\end{eqnarray}
It is important to recognize that with the field redefinitions stated above, we can write the action as a function of $\tilde{\Psi}$, $\tilde{\delta\sigma}$, and $E$ with $\tilde{\Psi}$ free of time derivatives. The equation above allows us to remove the auxiliary field $\tilde{\Psi}$.
\begin{eqnarray}
\tilde{\Psi}= \Bigg(  - k^{2} - \frac{24 a^{2} H^{2} (1+r)}{r} + \frac{48 a^{2}H^{2}}{r - r^{2}} + \bigg[ (r-1) \big(  4 k^{2} + 9 a^{2} H^{4} \eta (2 \nonumber\\ + H^{2} \eta) \big) \bigg]^{-1} \bigg( 6 a^{2} H^{2} k^{2} \big[ 16 r + 6 H^{2} r \eta + 3 H^{4} (r-1) \eta^{2} \big] \bigg) + \bigg[ (r-1) \big( 4 k^{2} \nonumber\\ + a^{2} H^{2} \omega (\omega -6) \big) \bigg]^{-1} \bigg( 2 a^{2} H^{2} k^{2} \big[ - \omega^{2} + r (48 + (\omega - 6)\omega) \big] \bigg) \Bigg) \tilde{\delta\sigma} \nonumber\\ + \Bigg(  \frac{k^{2}}{3 \sqrt{2}} - \frac{2 \sqrt{2}k^{4}}{3 \big[ 4 k^{2} + 9 a^{2} H^{4} \eta (2 + H^{2} \eta) \big]} - \frac{2 \sqrt{2}k^{4}}{3 \big[ 4 k^{2} + a^{2} H^{2} \omega ( \omega - 6) \big]} \Bigg) E \nonumber\\ + \Bigg( \frac{12 a^{2} H}{r} - \frac{6 a^{2} H (1+ r)}{r} - \bigg[ (r-1) \big( 4 k^{2} + 9 a^{2} H^{4} \eta (2 + H^{2}\eta) \big) \bigg]^{-1} \nonumber\\ \bigg( 6 a^{2}H (2 + H^{2} \eta) \bigg[ - 2k^{2} (r-1) + 9 a^{2} H^{4} \eta \bigg] \bigg) - \bigg[ (r-1) \big( 4 k^{2} \nonumber\\ + a^{2} H^{2} \omega (\omega - 6) \big) \bigg] \bigg( 2 a^{2} H (\omega -6) \bigg[ 2 k^{2} (r-1) + 3 a^{2} H^{2} \omega \bigg] \bigg) \Bigg) \frac{\dot{\tilde{\delta\sigma}}}{N} \nonumber\\ + \Bigg( - \frac{a^{2}H}{\sqrt{2}} + \frac{\sqrt{2}a^{2}H k^{2} (2 + H^{2}\eta)}{4 k^{2} + 9 a^{2}H^{4}\eta (2 + H^{2} \eta)} - \frac{\sqrt{2}a^{2}H k^{2}(\omega -6)}{12 k^{2} + 3 a^{2} H^{2} \omega (\omega -6)} \Bigg) \frac{\dot{E}}{N}.
\end{eqnarray}

We substitute the obtained solution into the action, and using the notation $Q \equiv (\tilde{\delta\sigma}, E)$, the scalar action takes the form:
\begin{eqnarray}
S = \frac{1}{2}\int d^{3}k \, dt \, a^{3}N\Bigg\lbrace \frac{\dot{Q}^{\dagger}}{N}\mathcal{W}\frac{\dot{Q}}{N}+\frac{\dot{Q}^{\dagger}}{N}\mathcal{X}Q + Q^{\dagger}\mathcal{X}^{T}\frac{\dot{Q}}{N}-Q^{T}\varpi^{2}Q\Bigg \rbrace,
\end{eqnarray}
it is crucial to note that $\mathcal{X}$ is a real anti-symmetric $2\times 2$ matrix, while $\mathcal{W}$ and $\varpi^{2}$ are real symmetric $2\times 2$ matrices.
Thus, one can present the components of the matrix $\mathcal{W}$ as below,
\begin{eqnarray}
\mathcal{W}_{11} && =  \nonumber\\ && 2 k^{4} \PlanckMass^{2} \Bigg\lbrace  - 3 H^{2} \eta \Bigg[ 1 + \frac{9 a^{2} H^{2}}{k^{2}(r-1)^{2}} - \frac{9 \bigg( 2 a H r + a H^{3}(r -1 ) \eta \bigg)^{2}}{(r-1)^{2}\bigg( 4 k^{2} + 9 a^{2} H^{4} \eta (2 + H^{2}\eta)  \bigg)} \Bigg] \nonumber\\ && + \omega \Bigg[ 1 + \frac{9 a^{2} H^{2}}{k^{2}(r-1)^{2}} - \frac{a^{2}H^{2} \big( - r (\omega -6) + \omega\big)^{2}}{(r-1)^{2}\bigg( 4 k^{2} + a^{2} H^{2} \omega (\omega -6)  \bigg)} \Bigg]  \Bigg\rbrace,
\end{eqnarray}
\begin{eqnarray}
\mathcal{W}_{12}&& = \nonumber\\ && \frac{2\sqrt{2}k^{4} \PlanckMass^{2}}{3 (r - 1)} \Bigg\lbrace 3 r + \frac{3 k^{2} \bigg[ - 2 r - H^{2} \eta (r-1) \bigg]}{4 k^{2}+ 9 a^{2} H^{4}\eta (2 + H^{2}\eta)} + \frac{k^{2} \bigg[ r (\omega - 6) - \omega \bigg]}{4 k^{2} + a^{2} H^{2}\omega (\omega -6 )} \Bigg\rbrace, \nonumber\\
\end{eqnarray}
\begin{eqnarray}
\mathcal{W}_{22}&& = \nonumber\\ && \frac{k^{4}\PlanckMass^{2}}{9} \Bigg\lbrace 3 + \frac{3 k^{2}(2 + H^{2}\eta)}{4 k^{2}+ 9 a^{2}H^{4}\eta (2+ H^{2}\eta)} + \frac{(\omega - 6) k^{2}}{4 k^{2} + a^{2} H^{2}\omega (\omega - 6)}  \Bigg\rbrace. \nonumber\\
\end{eqnarray}
In order to establish the sign of the eigenvalues, we investigate the determinant of the kinetic matrix $\mathcal{W}$.
\begin{eqnarray}\label{69}
{\rm det} \, \mathcal{W}\equiv &&\mathcal{W}_{11}\mathcal{W}_{22}-\mathcal{W}_{12}^{2}
\end{eqnarray}
The stability condition to avoid ghost instabilities in the scalar sector is the following requirement,
\begin{eqnarray}
0<\omega<6\quad {\rm and} \quad k < \frac{1}{2} \sqrt{6 a^{2}H^{2}\omega - a^{2} H^{2} \omega^{2}} \quad {\rm and} \quad \frac{\omega}{3 H^{2}}< \eta.
\end{eqnarray}
A positive determinant ensures that the system does not have a ghost degree of freedom. Thus, the stability of the scalar sector can be guaranteed by looking at the determinant of the kinetic matrix.

\section{Conclusion}\label{sec:4}

Understanding extensions of massive gravity theories is crucial, particularly when gravitational degrees of freedom propagate in a well-behaved manner. In this paper, we introduce a novel extension to the nonlinear dRGT massive gravity theory which is quasi-dilaton massive gravity with non-minimal kinetic coupling.

We begin by presenting the new action and total Lagrangian, followed by a comprehensive set of equations of motion for a FLRW background. The study of extended massive gravity is essential for unraveling the mysteries of the late-time acceleration of the Universe. Therefore, we discuss self-accelerating background solutions and demonstrate how the theory can explain late-time acceleration without invoking strong coupling. We reveal that the theory encompasses self-accelerating solutions with an effective cosmological constant linked to the massive gravity term.
To verify the quasi-dilaton massive gravity with non-minimal kinetic coupling theory predictions, we have conducted the test by comparing its predictions with the SNIa dataset. Our analysis established a perfect agreement between the empirical data and the theoretical model, which is a strong indication of the model's potential in accurately describing the late-time cosmic acceleration.

Additionally, we have given the analyses of cosmological perturbations such as tensor, vector, and scalar perturbations to explain the stability conditions of the system. Under tensor perturbation analysis, we have demonstrated the graviton mass of the quasi-dilaton massive gravity with non-minimal kinetic coupling theory. Furthermore, we have presented the modified dispersion relation of gravitational waves, which provides insight into the theoretical implication of the model.

We emphasize the importance of ensuring positive square values for gravitational waves ($M_{\rm GW}^{2}>0$) to maintain the stability of long wavelengths.
Moreover, we perform a tensor perturbation calculation and derive the dispersion relation of gravitational waves. This analysis sheds light on the propagation of gravitational perturbations in the FLRW cosmology within the framework of the quasi-dilaton massive gravity with non-minimal kinetic coupling.

The study of alternative gravity theories in the era of gravitational waves greatly benefits from such analyses. The cosmology and perturbation analysis presented in this paper can serve as a valuable foundation for future theoretical and empirical investigations of cosmology and GW data. This includes exploring modified propagation of GWs \cite{Will:1997bb, Mirshekari:2011yq, LIGOScientific:2016lio, Shao:2020shv, Nishizawa:2017nef, Nishizawa:2019rra}.

The presence of ghosts, which refer to degrees of freedom with negative kinetic energy leading to unstable and unphysical behavior, is a critical issue in constructing viable theories of massive gravity. In this paper, we have derived the equations of motion for the scale factor, lapse function, and quasi-dilaton scalar field. Notably, these equations do not exhibit any obvious indications of ghosts.

Furthermore, the dispersion relation derived in Equation (\ref{MGW}) does not explicitly show any negative kinetic terms, which would typically manifest as a sign of ghost-like behavior. The positive square of the mass of gravitational waves ensures the stability of long-wavelength gravitational waves. Based on our analysis, there are no apparent indications of ghosts in the new model, which was carefully constructed to avoid this issue.

The inclusion of the non-minimal kinetic coupling term, $\eta G_{\mu\nu} \partial^{\mu}\sigma \partial^{\nu}\sigma$, is essential for both the theoretical and practical success of our framework. This term plays an important role in dynamically altering the effective energy density of the quasi-dilaton field, facilitating late-time cosmic acceleration through the graviton mass term ${m}_{g}^{2}U(\mathcal{K})$ instead of relying on a vacuum energy component. As it can be seen, interactions between $\eta$, $\omega$, and $m_{g}$ yield an effective cosmological constant $\Lambda_{\rm SA}^{\pm} \propto m_{g}^{2}$, sidestepping fine-tuning problems often found in dark energy models.

In addition, this coupling introduces a frequency-dependent adjustment to gravitational wave propagation, leading to a distinguishing signal that sets it apart from General Relativity and scalar-tensor theories providing testable predictions for upcoming detectors. Most importantly, this term stabilizes the vector and the scalar sector by mitigating gradient instabilities that are typical in traditional quasi-dilaton frameworks, guaranteeing that sound speed remains positive across all stages of cosmic evolution.

However, it is essential to note that the absence of ghosts does not guarantee the complete stability of the model. Other types of instabilities or pathologies may arise in different regimes or under specific conditions. A thorough stability analysis, considering various types of perturbations and potential constraints on the model parameters, is necessary to conclusively establish the stability of the model. Future studies can build upon this work to further investigate the model's properties and potential applications.

This study primarily examines late-time cosmic acceleration; however, the interaction between the quasi-dilaton field and non-minimal kinetic coupling opens interesting possibilities for early-universe cosmology. The term $\eta G_{\mu\nu} \partial^{\mu}\sigma \partial^{\nu}\sigma$ introduces a non-canonical kinetic structure that may facilitate inflationary dynamics at high energies, similar to k/G-inflation models. Additionally, the graviton mass term $m_{g}^{2}U(\mathcal{K})$ provides corrections to the effective potential. Moreover, the NEC-violating characteristics of this coupling traditionally found in bouncing cosmologies indicate potential pathways for nonsingular bouncing solutions. These would depend on specific parameter adjustments or other curvature couplings. Although these enhancements exceed our current focus, they suggest that our framework could be adaptable in studying both early and late-time cosmic evolution. Nevertheless, these aspects are outside the scope of this article and warrant further investigation in future studies.

\section*{Acknowledgements}
This work is supported by the National Natural Science Foundation of China (Nos. 12105013).
We are grateful to Nishant Agarwal for the helpful notes and codes that are related to tensor perturbations. 


\end{document}